\documentclass[11pt]{article}
\usepackage{times}
\usepackage{amsmath}
\usepackage[margin=1in]{geometry}
\usepackage{lineno,hyperref}
\usepackage[table,x11names,dvipsnames,table]{xcolor}
\usepackage{authblk}
\usepackage{subcaption,booktabs}
\usepackage{graphicx}
\usepackage{multirow}
\usepackage{multicol}
\usepackage{placeins}
\usepackage[nolist,nohyperlinks]{acronym}
\usepackage[superscript]{cite}
\usepackage{tabularx}
\usepackage{float}
\usepackage{lipsum}
\usepackage[group-separator={,}]{siunitx}
\usepackage{geometry}
\usepackage{array}

\makeatletter
\def\@maketitle{
\centering 
\newpage
  \noindent
  \vspace{0cm}
  \let \footnote \thanks
    {\huge \textbf{{\@title}} \par} 
    \vskip 2em
    {\large
      \lineskip .5em
      \begin{tabular}[t]{c} 
        \@author
      \end{tabular}\par}
    \vskip 1em
  \par
  \vskip 1.5em
  }
\makeatother

\begin{document}

\title{The Energy Spectrum of Kaon from Lattice QCD}

\setlength{\affilsep}{0pt}
\author{
Arunangshu Bora\textsuperscript{1}, 
Anirban Mandal\textsuperscript{2}, 
Shreya Mittal\textsuperscript{3}, 
Mihir N Pandey\textsuperscript{4}, 
Rohan Rana\textsuperscript{5}, 
Harsh Saxena\textsuperscript{6}, 
Priyajit Jana\textsuperscript{7}, 
Rijul Dhumane\textsuperscript{8}, 
Tanmoy Bhowmik\textsuperscript{9}, and
}

\affil{Mentor: Dr. Martha Constantinou \\ 
Temple University, 1801 N Broad St, Philadelphia, PA 19122, USA\textsuperscript{1}}

\affil{\textsuperscript{1}\small \textit{BS-MS Physics Major (Passed) at Indian Institute of Science Education and Research (IISER), Tirupati}} 
\affil{\textsuperscript{2}\small \textit{BSc Physics (hons.), Calcutta University}} 
\affil{\textsuperscript{2}\small \textit{BS Data Science,Indian Institute of Technology Madras}}
\affil{\textsuperscript{3}\small \textit{BS-MS Physics Major at Indian Institute of Science Education and Research (IISER), Mohali}} 
\affil{\textsuperscript{4}\small \textit{BSc Physics (hons.), Kirori Mal College, University of Delhi}} 
\affil{\textsuperscript{5}\small \textit{BS-MS Physics Major at Indian Institute of Science Education and Research (IISER), Kolkata}} 
\affil{\textsuperscript{6}\small \textit{MS Physical Science at Jawaharlal Nehru Centre for Advanced Scientific Research, Bangalore}} 
\affil{\textsuperscript{7}\small \textit{BS-MS Physics Major at Indian Institute of Science Education and Research (IISER), Kolkata}} 
\affil{\textsuperscript{8}\small \textit{BSc(Hon.s) Computational Physics at Maharashtra Institute of Technology - WPU, Pune}} 
\affil{\textsuperscript{9}\small \textit{BSc Physics at Shahjalal University of Science and Technology, Sylhet}}

\date{}
\maketitle

\begin{abstract}
 This study presents the analysis of data related to the two-point function of kaon generated from
 lattice QCD simulations. Using gauge configurations of twisted-mass fermions, we obtain the correlation
 functions for 6 values of momentum for the kaon between 0 and 2GeV(both inclusive), we use statistical
 techniques such as jackknife resampling to derive the energy of the particle. The lattice results are
 compared to the continuum dispersion relation for the particle, to assess systematic uncertainties in
 the lattice data. We establish consistency with theory by comparing our results with the theoretical
 predictions.

\end{abstract}
\section{INTRODUCTION}
The strong force is one of the four fundamental forces in nature and is described by the theory of Quantum Chromodynamics (QCD). Gluons are the force carriers, and quarks can interact via this force. Kaons are among the many hadrons that arise from the strong force. They are further classified as mesons (particles made of a quark and anti-quark pair). They contain one strange quark (or antiquark) and an up- or down quark (or antiquark), which leads to four types of kaons.\\
We cannot use our perturbative tools to study the structure of hadrons due to the large strong coupling of the force in the low energy region (confinement). Another complication is the fact there are infinite degrees of freedom, that is quarks and gluons. These degrees of freedom manifest themselves physically in the QCD vacuum that consists of sea quarks and gluons. We define a finite volume hypercubic grid of space-time points (lattice) [\cite{gross2022}]. Since the lattice is defined in a finite volume, there are finite degrees of freedom, but, nevertheless, are of the order of billions.

The Quantum Chromodynamics (QCD) Lagrangian can be written as:

\begin{equation}
\mathcal{L}_{QCD} = -\frac{1}{4} F_{\mu \nu}^A F^{A \mu \nu} + i \bar{q} \gamma^\mu \left( \partial_\mu + i g_s \frac{1}{2} \lambda^A \mathcal{A}_\mu^A \right) q - \bar{q}_{\mathrm{R}} \mathcal{M} q_{\mathrm{L}} - \bar{q}_{\mathrm{L}} \mathcal{M}^{\dagger} q_{\mathrm{R}} - \theta \omega.
\end{equation}

The gluons are described by the gauge field $\mathcal{A}_\mu^A$, which belongs to the color group $\mathrm{SU}_c(3)$ and $g_s$ is the corresponding coupling constant. $F_{\mu \nu}^A$ is the field strength tensor.

One of the key points that differs QCD from Quantum Electrodynamics (QED) is that gauge invariance implies that the Lagrangian contains terms involving three or four gluon fields: in contrast to the photons, which interact among themselves only via the exchange of charged particles, the gluons would interact even if quarks did not exist [\cite{gross2022}].

To convert to the lattice QGT formulation, we introduce the spacing `$a$' between the points of a hypercubic grid, so the lattice field $n$ is defined on a discrete set of points $n = (n_1, n_2, n_3, n_4)$ in $R^4$ and $x = a n$. On a discrete lattice, the derivatives can be replaced by finite difference methods. Combining a rotation to Euclidean space and discretization, we can write the path integral as:
\begin{equation}
Z = \int \prod_i \mathrm{d} \phi_i \exp \left[ -S_E \left( \left\{ \phi_i \right\} \right) \right]
\end{equation}
where $\phi$ is the field at each site, and, $S_E$ is,
\begin{equation}
S_E = \int d^3 x \, d\tau \left[ -\frac{1}{4} F_{\mu \nu}^A F^{A \mu \nu} + i \bar{q} \gamma^\mu \left( \partial_\mu + i g_s \frac{1}{2} \lambda^A \mathcal{A}_\mu^A \right) q - \bar{q}_{\mathrm{R}} \mathcal{M} q_{\mathrm{L}} - \bar{q}_{\mathrm{L}} \mathcal{M}^{\dagger} q_{\mathrm{R}} - \theta \omega \right]
\end{equation}
The field is defined on a 4-D Euclidean domain, and we can observe that the partition function looks like that of a statistical mechanical system. This allows us to use Monte Carlo sampling techniques. The observables are calculated as expectation values. Their values change with lattice spacing, but they are usually related to our physical units by simple mathematical relations.

\section{METHODOLOGY}
We shall now pursue Lattice QCD more descriptively. It should be noted that the lattice description of Quantum field theories allows us to regularize them without the use of a perturbative approach, the lattice spacing $a$ acts as a cutoff for spacetime separation and equivalently $\frac{\hbar}{a}$ for momentum.[\cite{hari1985}] However, this lattice has no direct physical correspondence of its own at a finite $a$; we are eventually interested in obtaining observables at the continuum limit of the theory i.e., $a \rightarrow 0$.
\\
However, we have not encountered any explicit dependence on lattice spacing-$a$ in the theory; this problem is readily solved by looking at an analogous lattice description of a classical statistical system. This tells us that such a lattice formulation requires certain physical parameters analogous to Temperature $(T)$ in the classical regime, which allows us to determine all the measurable in terms of other physical parameters.
\\
\textbf{How - What - Why ?} \\
The structure of lattice action projects the physical picture of the system onto the lattice, such simulation scheme demands that quarks (or antiquarks) are placed at the lattice points which are linked through the gluonic field. The usual calculation of mass $(\vec{p}=0)$ for such a system requires the creation and annihilation of particle (kaon) inside the hypercubic lattice at time $t$ and $t^{\prime}$, the best way to extract the energy of the particle for a certain momentum state is through using Two-point correlation function.
\[
\begin{aligned}
C_M^{2pt}(t, \vec{p}) &= \sum_{\vec{x}} J_M(t, \vec{x}) J_M^{\dagger}(t_i, \vec{x_i}) e^{-i \vec{p} \cdot \vec{x}} \\
E_{eff}(t_j) &= \ln \left[ \frac{C^{2pt}(t_j)}{C^{2pt}(t_{j+1})} \right]
\end{aligned}
\]
The calculation of energy of the particle is done on a series of randomly generated configurations of QCD and extracting $C^{2pt}$ on them. The lattice we have used in this project has more than 2 billion lattice points.

{
\begin{center}
    \fbox{Lattice : (32, 32, 32, 64) \( \to 2 \times 10^{10} \) lattice points \( \to \) heavy computation}
\end{center}
}
\textbf{Jackknife resampling} is extensively used by the lattice practitioners, it is a ready-to-go tool used for error estimation and information gathering, using a small dataset. Which is what we deal with in lattice QCD systems. An alternative approach is the bootstrap analysis, which we do not discuss here. Regarding jackknife, the basic steps are:
\[
\begin{aligned}
f_m &= \sum_{m \neq n} \frac{f(x_n)}{N-1} \quad \langle f \rangle = \sum_m \frac{f_m}{N_{bin}}, \quad N_{bin} = N \\
\delta f &\equiv \sqrt{\frac{N-1}{N} \sum_m (f_m - \langle f \rangle)^2}
\end{aligned}
\]

\section{PROCEDURE}
We analyzed six values of the particle momentum, that is $0, 0.41, 0.83, 1.24, 1.66, 2.07 \mathrm{GeV}$.The statistics for each of these momenta is 1 198, 9 584, 9 584, 9 584, 28 752, 239 600, respectively. Also, in each case we have positive and negative values for the momentum. The procedure involves the following steps:

\begin{enumerate}
  \item \textbf{Mean value of 2pt functions for positive and negative momentum direction and Data Organisation:} We average the real part of two-point functions corresponding to the same magnitude of momentum in the z-direction at each time step. This includes averaging the functions for positive and negative z-direction while keeping the x and y directions constant. Then processed the files by separating the averaged real values of the two-point functions based on time. The data is then organized into a dictionary indexed by time values.
  \item \textbf{Data Averaging and time averaging:} Each Dictionary had 41 datasets. We averaged the values associated with same time value of datasets having same value of total momentum. Then using time reversal symmetry we averaged the 2pt values at time values (i) and (64-i) for i from 1 to 31. The values for time values 0 and 32 remain unchanged. This step is necessary to obtain a single plateau in the effective energy vs. time graph.
  \item \textbf{Jackknife Analysis:} We performed Jackknife resampling on the averaged data to calculate the Jackknife mean and standard deviation for each time step. This helps in estimating the uncertainty of the results.
  \item \textbf{Energy Calculation and Identification of Plateau:} We computed the ratio of two-point functions at consecutive time steps and took the natural logarithm to form the energy arrays. After plotting the $a E$ vs $t / a$ graph we observed a plateau-like region, selected the range in terms of time step $\left( [t_{\text{low}}, t_{\text{high}}] \right)$ and mark it as plateau region.
  \item \textbf{Calculation of Plateau Energy and Jackknife Error:} Then we calculated the final value of effective energy ($E_{\text{plateau}}$) at that particular value of momentum using the following formula:
  \[
  E_{\text{plateau}} = \frac{\sum_{t_i = t_{\text{low}}}^{t_{\text{high}}} E_{t_i} \left( \frac{1}{d t_i} \right)}{\sum_{t_i = t_{\text{low}}}^{t_{\text{high}}} \frac{1}{d t_i}}
  \]
  where  $E_{t_i}$  represents the energy at  $t_i$  and $d_i$ represents the jackknife standard deviation at that $t_i$. \newline We then calculate the Jackknife error and mean of the plateau fit values from the Jackknife binning results.
  \item \textbf{Comparison of Dispersion Relation:} Next we plot the dispersion relation using the results from various momentum analyses, applying Einstein's special theory of relativity. In natural units: $a^2 E^2 = a^2 m^2 + a^2 (p_x^2 + p_y^2 + p_z^2)$ where $a$ is the lattice unit.
\end{enumerate}

\section{RESULTS and CONCLUSION}
Following the already mentioned procedure, two representative plateaus are shown below for the momentum $0.83 \mathrm{GeV}$ and $1.24 \mathrm{GeV}$ (where the x and y component is 0) and momentum $\sqrt{6}$ (where the resultant of x and y component is $\sqrt{2}$).

\begin{figure}[h!]
    \begin{minipage}{0.5\textwidth}
        \centering
\includegraphics[width=\linewidth]{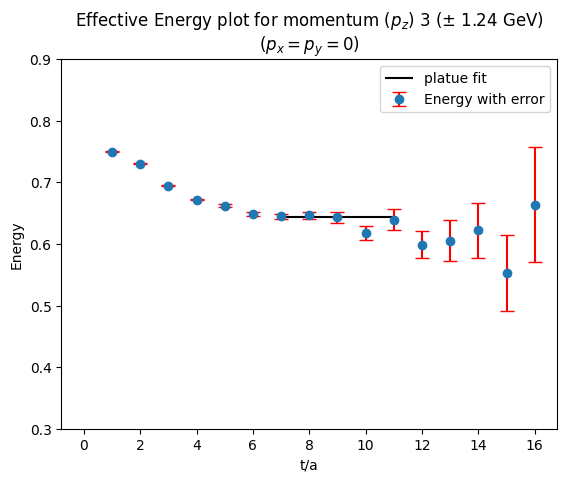}
        \caption*{ \textit{plateau} = 0.6430214 ± 0.0000028}
    \end{minipage}%
    \begin{minipage}{0.5\textwidth}
        \centering
        \includegraphics[width=\linewidth]{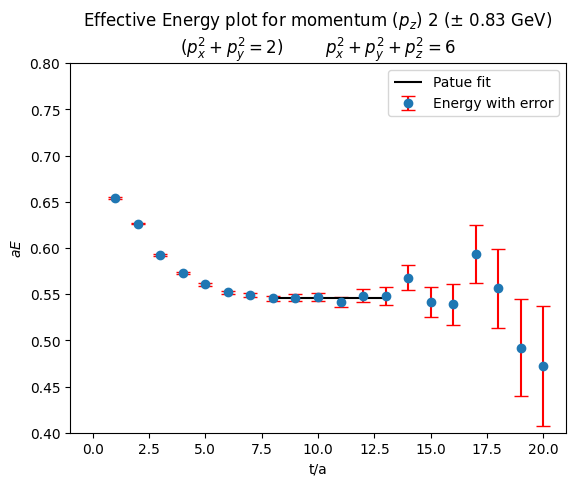}
        \caption*{ \textit{plateau} = 0.5456180 ± 0.0000017}
    \end{minipage}
\end{figure}

Plateau fitted values for different combinations of momenta boost are given below in the table: \newline
 Now we compare our computationally generated data (through Lattice QCD) with its theoretical counterpart using the dispersion relation

\begin{figure}[H]
    \centering
    \includegraphics[width=0.5\linewidth]{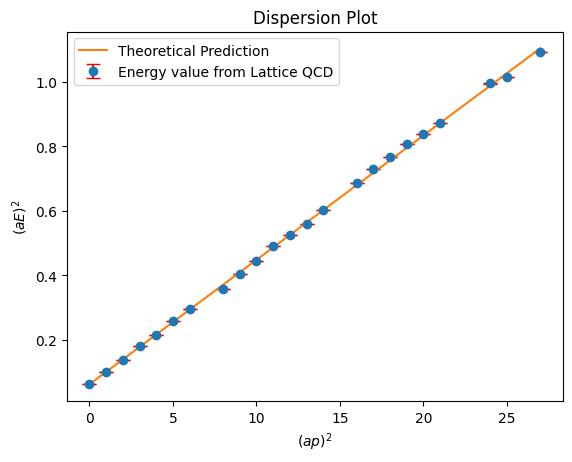}
    \caption*{}
    \label{}
\end{figure}

\begin{table}[h!]
\centering
\begin{tabular}{|c|c|c|}
\hline
Momentum Configuration $(p_x^2 + p_y^2 + p_z^2)$  & Effective  Energy $(aE)$ & Error \\
\hline
$0 \, (0 \, \text{GeV})$ & 0.2503897571908317 & $2.5791416304381206e-07$ \\
\hline
$1 \, (\pm 0.41 \, \text{GeV})$ & 0.31812734594752745 & $4.424495113154001e-07$ \\
\hline
2 &  0.3732724854969631 & $ 9.045264952465114e-07$ \\
\hline
3 & 0.4247528925508162 & $6.083625329961796e-07$ \\
\hline
$4 \, (\pm 0.83 \, \text{GeV})$ & 0.463223332509825 & $1.5474629322450727e-06$ \\
\hline
5 & 0.5085321794457005 & $ 1.4598611197100595e-06$ \\
\hline
6 &  0.5438130645474166 & $1.8307294007377191e-06$ \\
\hline
8 & 0.5980768744904814 & $6.925773793514575e-06$ \\
\hline
$9 \, (\pm 1.24 \, \text{GeV})$ & 0.6356445374137877 & $4.329590808434425e-06$ \\
\hline
10 & 0.6668746069233606 & $4.8090156941907785e-06$ \\
\hline
11 & 0.69971653807267 & $ 3.5826855165384203e-06$ \\
\hline
12 & 0.7251039340525522 & $6.6357252565823165e-06$ \\
\hline
13 & 0.747680677161704 & $ 9.214527586761378e-06$ \\
\hline
14 &  0.7759013508398236 & $ 6.838403679556357e-06$ \\
\hline
$16 \, (\pm 1.66 \, \text{GeV})$ &  0.8289985937907043 & $ 2.8623048888467666e-06$ \\
\hline
17 & 0.854196899649006 & $7.11910045851984e-06$ \\
\hline
18 & 0.8759581328257354 & $ 9.001944872741864e-06$ \\
\hline
19 & 0.8991794712094623 & $1.3042694690290442e-05$ \\
\hline
20 & 0.91539787219901 & $3.9442412257725426e-06$ \\
\hline
21 & 0.9344281530391869 & $4.464917258240312e-06$ \\
\hline
24 & 0.9972232821194545 & $9.119105857064124e-06$ \\
\hline
$25 \, (\pm 2.07 \, \text{GeV})$ & 1.00780980951629 & $2.303036367856831e-06$ \\
\hline
27 & 1.0456348117309648 & $2.2154319910003868e-06$ \\
\hline
\end{tabular}
\caption*{}
\end{table}

From the above plot we clearly see that our data clearly agrees with its theoretical counterpart so we can say that Lattice QCD can be one of the promising approach to probe the kaons in lower as well as for higher energy spectrum.

All the codes used for the analysis are given here: \small \url{http://bit.ly/3Xsvume}

\section{ACKNOWLEDGEMENTS}
We thank Dr. Martha Constantinou for mentoring us throughout the project and the REYES program for allowing us to work on this project. We also thank Josh Miller and Isaac Anderson for their advice on technical aspects. M.C. acknowledges financial support from the U.S. Department of Energy, Office of Nuclear Physics, Early Career Award under Grant No. DE-SC0020405, and Grant No. DE-SC0025218. Interactions within the Quark-Gluon Tomography (QGT) Topical Collaboration, funded by the U.S. Department of Energy, Office of Science, Office of Nuclear Physics, with Award DE-SC0023646, have benefited this work. This research used resources of the National Energy Research Scientific Computing Center, a DOE Office of Science User Facility supported by the Office of Science of the U.S. Department of Energy under Contract No. DE-AC02-05CH11231 using NERSC award NP-ERCAP0027642, as well as NERSC award ALCC-ERCAP0030652. Also, some of the data used in this work have been produced using a computational award from Oak Ridge Leadership Computing Facility through the Summit Plus program (project: NPH160).

\end{document}